# Fabrication of highly dense isotropic Nd-Fe-B bonded magnets via extrusion-based additive manufacturing

**Ling Li[1], Kodey Jones[1], Brian Sales[1], Jason L. Pries[1], I. C. Nlebedim[2], Ke Jin[1], Hongbin Bei[1], Brian Post[1], Michael Kesler[1], Orlando Rios[1], Vlastimil Kunc[1], Robert Fredette[3], John Ormerod[3], Aaron Williams[4], Thomas A. Lograsso[2], and M. Parans Paranthaman[1*]**

[1]*Oak Ridge National Laboratory, Oak Ridge, TN 37831, USA*

[2]*Ames Laboratory, Ames, IA 50011, USA*

[3]*Magnet Applications, Inc., DuBois, PA 15801, USA*

[4]*Arnold Magnetics Technologies, Rochester, NY 14625, USA*

**Abstract**

Isotropic bonded magnets with a high loading fraction of 70 vol.% Nd-Fe-B are fabricated via an extrusion-based additive manufacturing, or 3D printing system that enables rapid production of large parts for the first time. The density of the printed magnet is 5.15 g/cm$^3$. The room temperature magnetic properties are: intrinsic coercivity $H_{ci}$ = 8.9 kOe (708.2 kA/m), remanence $B_r$ = 5.8 kG (0.58 Tesla), and energy product $(BH)_{max}$ = 7.3 MGOe (58.1 kJ/m$^3$). The as-printed magnets are then coated with two types of polymers, both of which improve the thermal stability at 127 °C as revealed by flux aging loss measurements. Tensile tests performed at 25 °C and 100 °C show that the ultimate tensile stress (UTS) increases with increasing loading fraction of the magnet powder, and decreases with increasing temperature. AC magnetic susceptibility and resistivity measurements show that the 3D printed Nd-Fe-B bonded magnets exhibit extremely low eddy current loss and high resistivity. Finally, we show that through back electromotive force measurements that motors installed with 3D printed Nd-Fe-B magnets exhibit similar performance as compared to those installed with sintered ferrites.

Permanent magnet (PM) drives are the heart of modern electric and hybrid electric vehicle technologies. As the strongest PM, $Nd_2Fe_{14}B$ based magnets (Nd-Fe-B) with a $BH_{max}$ of up to 64 MGOe have been the focus of intense research in terms of both basic manufacturing science and engineering since their discovery in 1984 [1]. Nd-Fe-B magnets can be manufactured in several forms including sintered and bonded. Polymer bonded magnets have gained rising importance due to their growing applications ranging from hard disk drives to electric motors where shape design flexibility and high energy product are simultaneously required [2,3]. Conventionally, Nd-Fe-B isotropic bonded magnets are fabricated by injection molding and compression bonding, which have a typical magnet powder loading fraction of up to 65 and 80 vol.%, respectively. Both of the above methods can manufacture magnets to near net shape with appropriate tool design [4]. As the maximum energy product of fully dense Nd-Fe-B magnets has almost saturated, progress on improving processing methods is crucial to produce low-cost magnets with superior magnetic and mechanical properties as well as improved thermal stability [5,6].

Recently, research on the manufacturing of Nd-Fe-B rare earth magnets has experienced renewed interest because of innovative developments in additive manufacturing (AM) [7-12]. The primary advantages of AM lie in the ability to minimize waste, eliminate tooling requirement, and realize arbitrary size and shape design. In particular, the merit of minimum waste could greatly reduce the production cost and ease the rare earth supply crisis as Nd-Fe-B is composed of rare earth elements such as Dy and Nd, which are identified as critical materials by the U.S. Department of Energy [13]. Direct metal printing of metallic Nd-Fe-B magnets is demonstrated in Jacimovic *et al.*'s promising work using a Selective Laser Melting method [11]. Despite the complex ternary phase diagram of the Nd-Fe-B system, high coercivity is achieved in the final printed part, which is ascribed to non-equilibrium effects that occur during the cooling of the melts from high temperatures. Nevertheless, the porosity in the printed magnets still needs to be optimized to enhance the energy product. On the other hand, nylon bonded Nd-Fe-B magnets with a loading fraction of 54 vol.% have been fabricated by Huber *et al.* [8] utilizing a commercial end-user 3D printer, which uses extruded filaments as the feedstock materials. The advantage of this method is high precision due to the thin layer height of 0.1 mm, which makes it well-suited to produce small-sized magnets used in office appliances, *etc*.. Recently, we have demonstrated a different extrusion-based 3D printing method to fabricate nylon bonded Nd-Fe-B magnets with a loading fraction of 65 vol.% [9]. Conventional AM systems are typically limited to small sizes and slow rates; the Big Area Additive Manufacturing system (BAAM) developed at Oak Ridge National Laboratory, in contrast, can produce very large parts at very high rates using cost effective feedstock materials (magnet nylon compounded pellets). BAAM magnets with 65 vol.% loading fraction exhibit lower porosity (8%) and superior magnetic and mechanical properties [9].

To compete with injection and compression molding, which are the established methods to fabricate complex-shaped isotropic bonded magnets, we aim to increase the loading fraction of the magnet powder in the nylon binder in our 3D printing process and thereby increase the magnetic strength. In this study, we have successfully printed 70 vol.% nylon bonded Nd-Fe-B magnets in various shapes (see supplementary material, Fig. S1). In addition, we have also coated the as-printed magnets with two types of epoxy-based polymers to improve their thermal stability. The microstructure, magnetic and mechanical properties, flux aging loss, and eddy current loss are reported. We also successfully demonstrated the performance of the 3D printed magnets in a DC motor configuration.

Computer-aided design (CAD) software was used as the first step of the 3D printing manufacturing process. The Big Area Additive Manufacturing (BAAM) printer (schematically shown in Ref. [9]) used similar methods of taking a CAD file to guide the print head for layer by layer growth during the printing process. The geometric designs were created and saved as standard triangle language (.STL) files, which were then "sliced" in the printer software which created the layers and path for each layer. Composite pellets made of 70 vol.% isotropic MQP B+ powder [14] and 30 vol.% polyamide provided by Magnet Applications *Inc.* were used as the feedstock materials for the BAAM system. The material was deposited at 255 °C onto a heated print bed kept at ~ 95 °C. Note that the preheating of the bed is to enhance the bonding between layers as the beads are bonded together primarily via thermal fusion which depends on the temperature, viscosity, and surface area [15]. The nozzle diameter was 0.3 inch, and the layer thickness was 0.15 inch. The printing speed was 1 inch/second. The nozzle was equipped with a z-tamping attachment to obtain a leveled surface, and to strengthen the bonding between layers. More details about the compounding process of the pellet feed and the print parameters have been reported previously [9].

The magnetic hysteresis loops were measured from 20 to 120 °C, using a closed loop magnetic hysteresisgraph measurement system at Arnold Magnetics *Inc.*. Cross-sections of the as-printed sample and fracture surface were examined by Scanning Electron Microscopy (Hitachi S-4700). Flux measurements were carried out using Helmholtz Coils with a fluxmeter (Model 2130) from Magnetic Instrumentation *Inc.*. The samples were magnetized at 9 Tesla and then aged at elevated temperatures. The as-printed samples were treated with two types of polymer coatings in an attempt to improve the thermal stability. The first type is a 3M ScotchWeld DP100 epoxy (Process I), and the second type is a high temperature silica ceramic VHT coating (Process II). Tensile tests were performed with a screw-driven tensile testing machine (Instron) at an engineering strain rate of $10^{-3}s^{-1}$ and temperatures of 20 and 100 °C, in ambient air. Dog bone shaped samples (see supplementary material, Fig. S2) with gauge dimension of 3.1x1.5x12.7 $mm^3$ were used in all testing. Because the plastic deformation is limited, we only determined the ultimate tensile stress, which is calculated by the maximum load divided by the original cross section area. Eddy current losses were estimated from AC susceptibility measurements. Resistivity measurements were done with a four-probe resistivity measurement setup. The 70 vol.% BAAM magnets were installed into a 12 V - 750 RPM - DC motor and evaluated.

Table 1 summarizes the magnetic characteristics of the BAAM printed 70 vol.% Nd-Fe-B bonded magnets with a density of ρ ~ 5.15 $g/cm^3$. The room temperature properties are $B_r$ = 5.8 kG (0.58 Tesla), $H_{ci}$ = 8.9 kOe (708.2 kA/m), $H_c$ = 4.8 kOe (382 kA/m) and $BH_{max}$ = 7.3 MGOe (58.1 $kJ/m^3$). According to Magnequench's website [14], the injection molded magnet made from MQP B+ powders has a density of ρ ~ 5.0 $g/cm^3$, and $BH_{max}$ = 6.2 MGOe (49.3 $kJ/m^3$). These reduced properties result from the limitation of the loading fraction of the injection molding process, typically 65 vol.%. The loading fraction limitation is because of the viscosity of the composite increases as the loading fraction increases, and this results in cavities in the injection molding process. The magnetic characteristics of the 70 vol.% BAAM magnets at elevated temperatures up to 120 °C are also presented in Table 1. It is not surprising that all the characteristics decrease with increasing temperatures. The magnetic performance with increasing temperature can be evaluated from the temperature coefficients of $B_r$, $H_{ci}$ and $BH_{max}$, which are, as shown in Table 2, -0.18%/°C, -0.33%/°C, and -0.36%/°C, respectively. These values are also close to the reported thermal coefficients for MQP B+ powders [4]. In fact, the primary drawback of Nd-Fe-B magnets is the unsatisfactory magnetic performance at elevated temperatures, which is related to the relatively high temperature coefficients of $B_r$ and $H_{ci}$, leading to a rapid deterioration in the intrinsic coercivity and flux density. Therefore, Sm-Co magnets are frequently adopted when high temperature operation is needed.

The thermal stability of magnets can also be compared by measuring the room temperature flux density of the samples before and after aging. The flux loss is related to a magnetization reversal mechanism occurring with rising temperature [6]. The total flux loss is composed of recoverable loss and irreversible loss, and the latter is due to the oxidation of the Nd-Fe-B powder [6]. The percentage of flux loss at a certain temperature is a direct reflection of the thermal stability of the magnet. The flux aging loss with time for the as-printed 70 vol.% BAAM magnets is presented in Fig. 1(a). Note that all the flux loss samples are rectangular shaped with a consistent permanence coefficient of ~2. Industry standards dictate that magnets losing more than 5% flux density are deemed unfit for use [16]. In this sense, the maximum operating temperature for the as-printed samples is ~ 110 °C. In general, the temperature coefficient and/or the thermal stability of the magnets depend on the powder type and manufacturing process such as compact pressure and temperature [17]. It is shown that epoxy bonded Nd-Fe-B magnets made from MQP B+ powders experienced 15% flux aging loss after being held at 180 °C for 100h, indicating that this powder is unsuitable for elevated temperature applications [4]. In general, powders with higher intrinsic coercivity $H_{ci}$ are more suitable for high temperature applications. With a given type of powder, however, the way to improve thermal stability is to prevent the magnets from being oxidized through the addition of a protective layer either on the individual powders or on the whole magnet. In this study, we coat the as-printed 70 vol.% BAAM magnets with two types of polymers (Process I and Process II), and compare the flux aging loss for the treated and as-printed samples, and determine the maximum operating temperatures. Process I is a 3M ScotchWeld DP100 two-part epoxy due to its affordability, industry prevalence, and temperature capabilities. Process II is the high temperature silica ceramic VHT coating. Fig. 1(b) shows the flux loss after aging at 127 °C for the as-printed, Process I, and Process II treated samples. The thermal stability is most improved by treating the as-printed samples with Process I. As shown in Fig. S3 (see supplementary material), the thermal stability at 102 °C is also improved with Process I coating.

| Temperature (°C) | $B_r$ (kG) | $H_{ci}$ (kOe) | $H_c$ (kOe) | $BH_{max}$ (MGOe) |
|---|---|---|---|---|
| 20 | 5.82 | 8.85 | 4.82 | 7.25 |
| 60 | 5.10 | 7.34 | 4.01 | 5.36 |
| 80 | 4.99 | 6.92 | 3.89 | 5.12 |
| 100 | 4.91 | 6.35 | 3.77 | 4.93 |
| 120 | 4.78 | 5.92 | 3.62 | 4.64 |
| Temp Coefficient (%) | -0.18 | -0.33 | -0.25 | -0.36 |

***Table 1:*** *Magnetic characteristics of the BAAM printed 70 vol.% isotropic Nd-Fe-B magnets at temperatures ranging from 20 to 120 °C.*

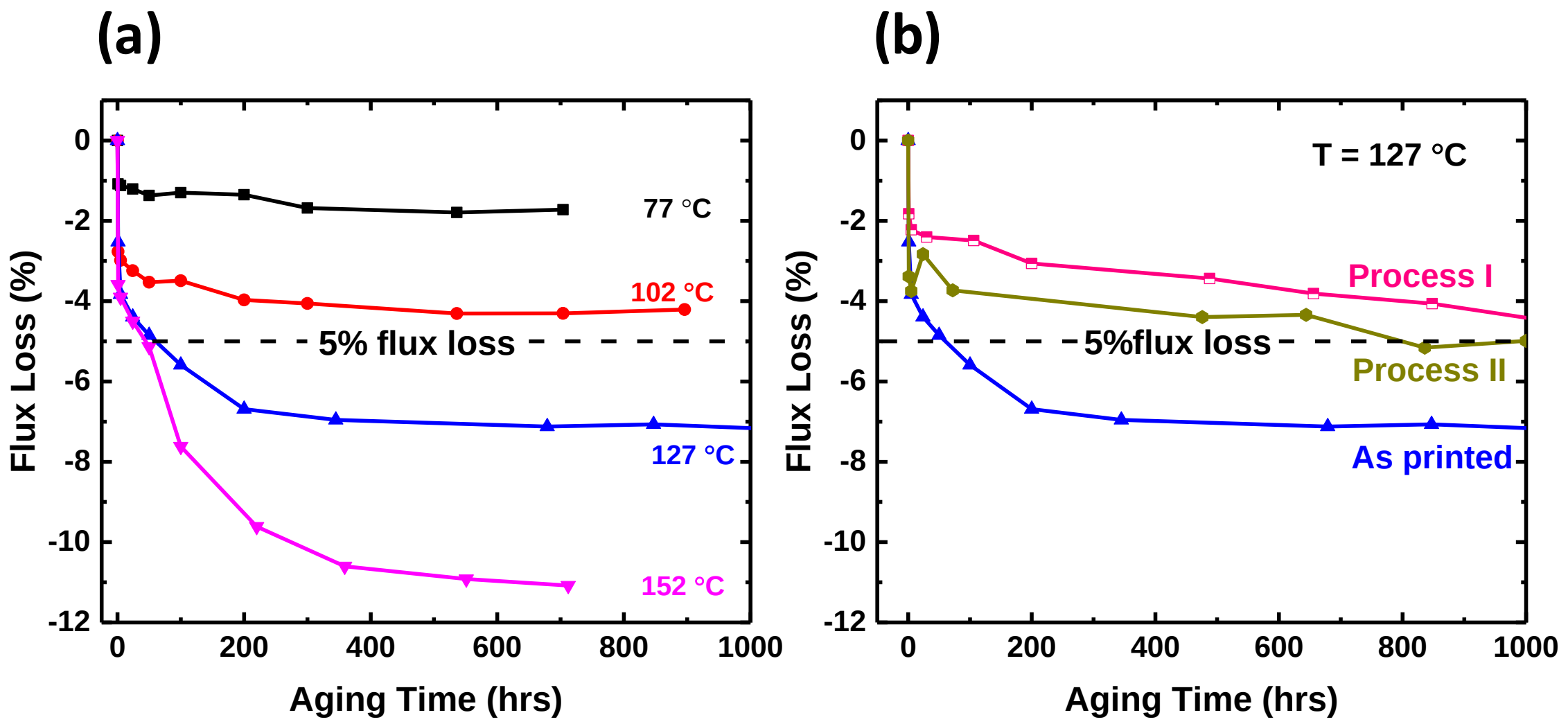

***Figure 1:*** *Flux aging loss for 70 vol.% BAAM magnets as a function of aging time (0–1000 hrs) for (a) as-printed samples at various temperatures; (b) as-printed and coated samples at 127 °C.*

*Note that all the samples are rectangular shaped with a consistent permanence coefficient ($P_C$) of 2.*

| Sample | Temperature (°C) | Ultimate Tensile Stress (MPa) | Tensile Strain (%) |
|---|---|---|---|
| 70 vol.% axial | 25 | 12.62 (0.16) | 2.14 (0.06) |
| | 100 | 5.95 (0.61) | 1.84 (0.16) |
| 70 vol.% transverse | 25 | 4.65 (0.95) | 1.00 (0.19) |
| | 100 | 3.00 (0.11) | 1.63 (0.16) |
| 65 vol.% axial | 25 | 9.90 (1.70) | 2.21 (0.39) |
| | 100 | 4.25 (0.09) | 3.80 (0.54) |

***Table 2.*** *Mechanical properties of the 70 vol.% and 65 vol.% BAAM magnets in both axial and transverse directions, and at 20 °C and 100 °C.*

Fig. 2 (a) presents an SEM image of the polished microstructure of a 70 vol.% BAAM magnet. The magnet particles appearing as bright areas are embedded in the Nylon-12 binders shown as black. The magnet particles have a flat plate-like morphology with dimensions ranging from several µm up to a hundred µm. This morphology of the isotropic particles results from melt-spun ribbon cut samples. These types of powders are also used to fabricate compression molded magnets [3].

Fig. 2(b) shows an SEM image of the fractured surface. The magnet particles are pulled out from the polymer binder. Table 2 presents the ultimate tensile stress (UTS) and strain for BAAM printed 70 vol.% magnets in both the axial (in layer plane) and transverse directions (perpendicular to layer plane) and 65 vol.% magnets in the axial direction. Four features are noteworthy: 1) The UTS and tensile strain of BAAM magnets are close to those of injection molded magnets, namely with a room temperature UTS of ~ 10 MPa and a strain of ~ 3%; 2) The UTS is significantly direction dependent, with the transverse direction being weaker than the axial direction. This should also apply to the 65 vol.% magnet even through its transverse direction is not measured; 3) The UTS increases with increasing loading fraction. Here the axial UTS increases from 9.90 to 12.62 MPa at 25 °C, and increases from 4.25 to 5.95 MPa at 100 °C when the loading is increased from 65 vol.% to 70 vol.%; 4) For a given loading fraction, the UTS decreases with increased temperature. Note that the latter two observations are consistent with the literature values [18]. Even though BAAM magnets have a much lower UTS compared to sintered magnets, they exhibit a higher degree of ductility, allowing for some plastic deformation.

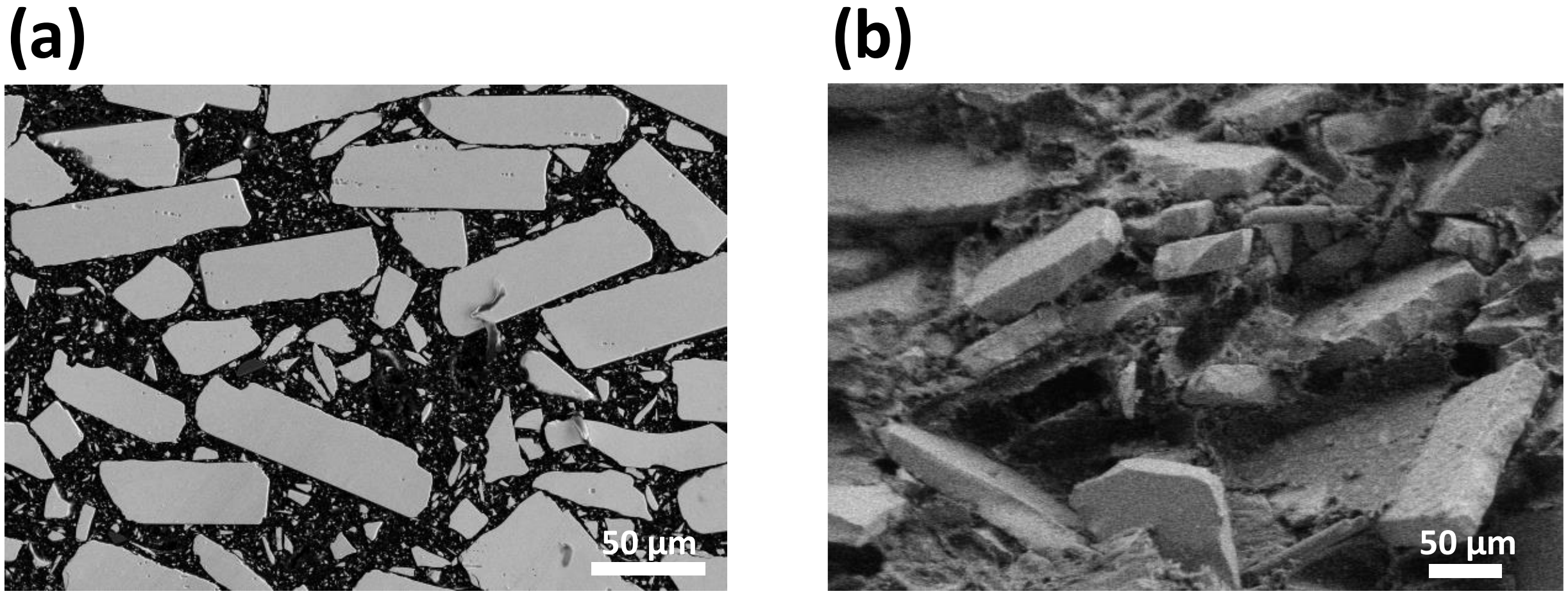

***Figure 2.*** *SEM images of the 70 vol.% BAAM magnets: (a) cross section of the as printed sample; (b) fractured surface.*

Even though sintered magnets offer high residual magnetic flux density which contributes to a high torque for a motor magnet, the high eddy current loss resulting from a high electrical conductivity leads to lower efficiency converting energies between magnetic and mechanical. In the present study, eddy current losses are estimated from AC magnetic susceptibility measurements where both the real (M') and imaginary (M'') parts are measured. The imaginary part of the susceptibility M'', corresponds to dissipative losses in the sample which can result in substantial heating. In electrical conductors, the dissipation is due to eddy currents, but in ferromagnets there are additional losses such as losses due to irreversible domain wall motion or hysteresis loss.

Figure 3 presents the total loss fraction as a function of AC magnetic field frequency (amplitude of 10 Oe) for an anisotropic Nd-Fe-B sintered magnet (for comparison) and BAAM printed 70 vol.% Nd-Fe-B magnets. Both magnets were demagnetized for this measurement. The measured DC electrical resistivity, ρ, of the bonded and sintered sample is 170 mΩ cm and 150 μΩ cm, respectively. Since the eddy current losses [19,20] are proportional to 1/ ρ, the bonded magnet will have significantly less eddy current heating. This is consistent with the loss fraction of the 3D printed sample which is extremely low with M''/M' well below 1% with increasing frequency, whereas the sintered sample exhibits a higher loss fraction of ~20%. In the simplest models the eddy current loss should increase as the square of the frequency. Although the losses for the sintered magnet increase with frequency, the increase is closer to linear than quadratic, indicating additional loss mechanisms or that the model is too simple. Even though the energy product of bonded magnets is sacrificed due to the incorporation of a non-magnetic polymer binder, the advantages of substantial design freedom and low eddy current loss thus high conversion efficiency will make 3D printed magnets rival sintered magnets for some motor applications. In fact, there exists a number of articles in the literature discussing the realization of PM materials reduction in motors through innovative designs [21-24]. For example, Wu *et al*. [24] has achieved rare earth reduction in a U-shaped hybrid permanent magnets assisted synchronous reluctance motor. Furthermore, 3D printed bonded magnets, with the advantages of shape flexibility and low eddy current loss, can be applied to an axial gap motor to achieve both high torque and high efficiency [25]. Besides the aforementioned considerations, weight reduction is also a key advantage of bonded magnets facilitating their adoption in the automobile industry.

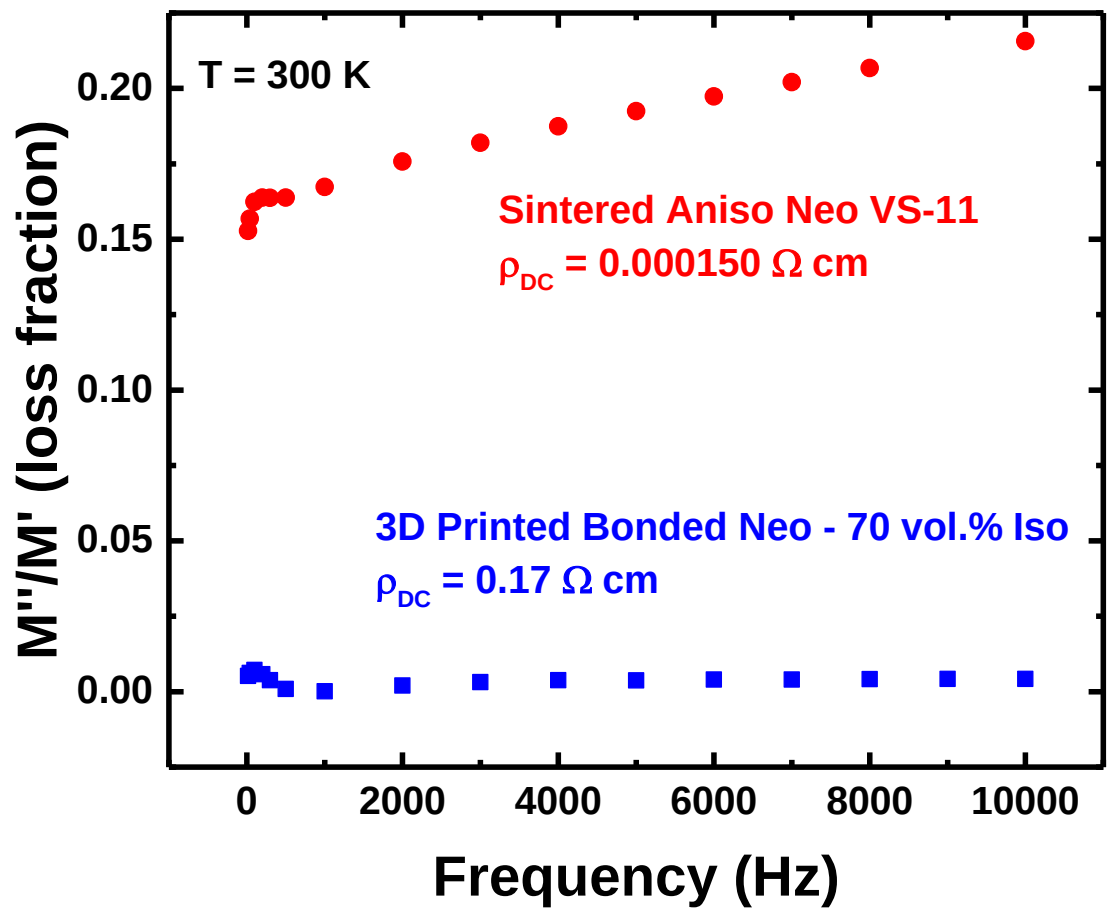

***Figure 3.*** *The eddy current loss fraction of 70 vol.% BAAM magnets and sintered anisotropic Neo VS-11magnets.*

To test the applicability of the 70 vol.% BAAM magnets for motor applications, we replaced the existing two arc segments of sintered ferrite magnet in a DC motor with BAAM magnets. Fig. S4 (see supplementary material) shows the images of the DC motor and the arc magnets with dimensions specified. A mounting piece was designed and 3D printed to securely hold the motors and keep the shafts concentric as shown in Fig. S5 (see supplementary material). The motors' back electromotive force (back-EMF) were tested in a back-to-back fashion as shown in the inset of Figure 4. A voltage was applied to the motor on the right hand side to drive the motor of the left, which is the motor under test. The average output voltage of the motor under test was measured. This voltage is proportional to the flux produced by the magnets and the speed of the rotor. Figure 4 shows the back-EMF test data for the motor installed with the original ferrite magnets and the BAAM magnets. Given the same magnet shape, the magnitude of EMF is a reflection of the strength of the permanent magnet. It can be seen from Figure 4 that the ferrite motor and BAAM magnet motor exhibit similar performance. The back-EMF constant (slope) of the line is 2.07V/p.u. (per unit values) for the ferrite magnet motor and 1.91V/p.u. for the printed magnet motor. It should be noted that the printed magnets are slightly smaller than the original ferrite magnets, which may account for much of the 8% drop in the back-EMF as the loaded flux density depends on the magnet strength (energy product) and magnet volume. Our study shows that BAAM magnets are suitable for small motor applications. To further improve the remanence $B_r$ and energy product $BH_{max}$ of 3D printed bonded magnets, a higher density (lower porosity) needs to be achieved. Optimally, in-situ magnetic alignment printing is needed to further enhance $B_r$ and thus $BH_{max}$ of the printed anisotropic magnet product. Even though this endeavor remains challenging, partly due to the complication in the mechanical design of the external magnetic field fixture to be attached to the 3D printer, efforts are underway.

The ability to create a near-net shaped magnet results in less post processing, less (nearly zero) waste generation, and a wide range of applications for a single manufacturing platform. In this study, nylon bonded Nd-Fe-B magnets with a high loading fraction of 70 vol.% are fabricated via an extrusion-based additive manufacturing process for the first time. The printed magnets exhibit superior magnetic properties, compared to injection molded magnets, while maintaining substantial geometry flexibility. Motors installed with the 3D printed magnets exhibits similar performance as compared to those installed with sintered ferrites. In addition, the extrusion-based additive manufacturing method via the BAAM system can be widely applied to net-shape manufacture other functional magnets like SmCo, SmFeN, $Fe_{16}N_2$, ferrites, and hybrids of more than one composition with binder materials like nylon, PPS (polyphenylene sulfide), etc..

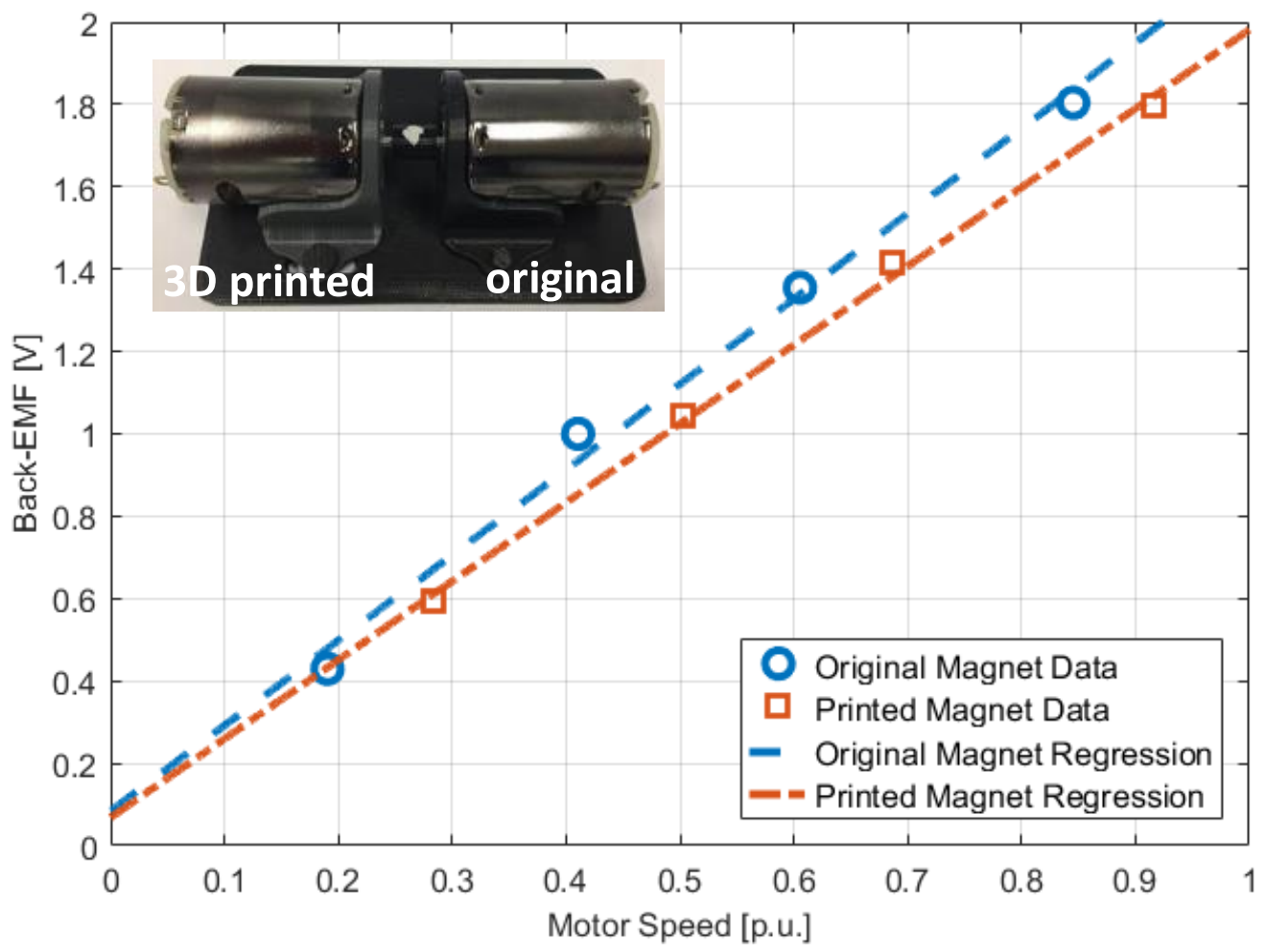

*Figure 4. Comparison of the back electromotive force (back - EMF) for motors installed with the original ferrites (right image of the inset) and the70 vol.% BAAM magnets (left image of the inset). The inset shows the back-to-back motor testing configuration.*

See supplementary material for the images of printed magnets and dog-bone design for tensile tests, flux aging loss studies of printed magnets at 102 °C, and images of the dc motor tested and motor mount.

This work was supported by the Critical Materials Institute, an Energy Innovation Hub funded by the U.S. Department of Energy, Office of Energy Efficiency and Renewable Energy, Advanced Manufacturing Office.

## Additional information

This manuscript has been authored by UT-Battelle, LLC under Contract No. DE-AC05-00OR22725 with the U.S. Department of Energy. The United States Government retains and the publisher, by accepting the article for publication, acknowledges that the United States Government retains a non-exclusive, paid-up, irrevocable, world-wide license to publish or reproduce the published form of this manuscript, or allow others to do so, for United States Government purposes. The Department of Energy will provide public access to these results of federally sponsored research in accordance with the DOE Public Access Plan (http://energy.gov/downloads/doe-public-access-plan). All the authors have no competing financial interests.

# Supplemental Information

# Fabrication of highly dense isotropic Nd-Fe-B bonded magnets via extrusion-based additive manufacturing

**Ling Li[1], Kodey Jones[1], Brian Sales[1], I. C. Nlebedim[2], Ke Jin[1], Hongbin Bei[1], Brian Post[1], Michael Kesler[1], Orlando Rios[1], Vlastimil Kunc[1], Robert Fredette[3], John Ormerod[3], Aaron Williams[4], Thomas A. Lograsso[2], and M. Parans Paranthaman[1*]**

[1]*Oak Ridge National Laboratory, Oak Ridge, TN 37831, USA*

[2]*Ames Laboratory, Ames, IA 50011, USA*

[3]*Magnet Applications, Inc., DuBois, PA 15801, USA*

[4]*Arnold Magnetics Technologies, Rochester, NY 14625, USA*

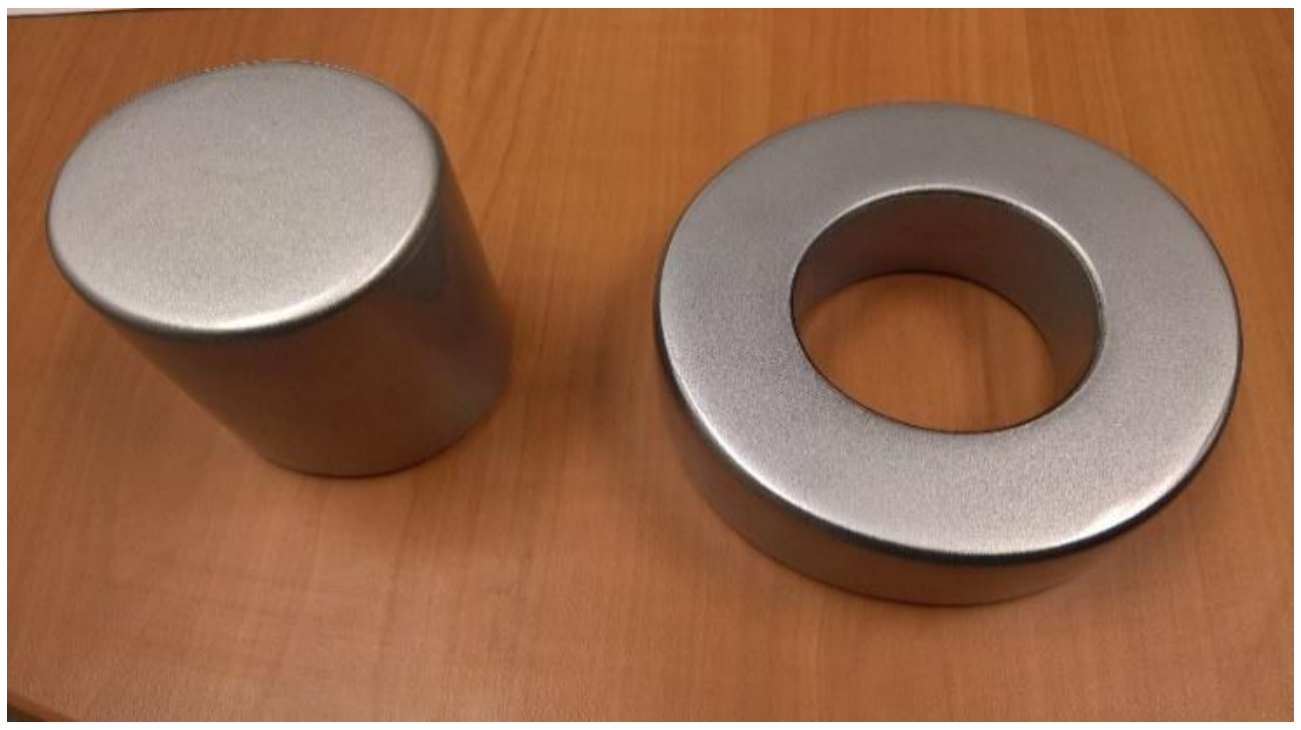

*Fig. S1. Image of the BAAM 70 vol.% Nd-Fe-B bonded magnets.*

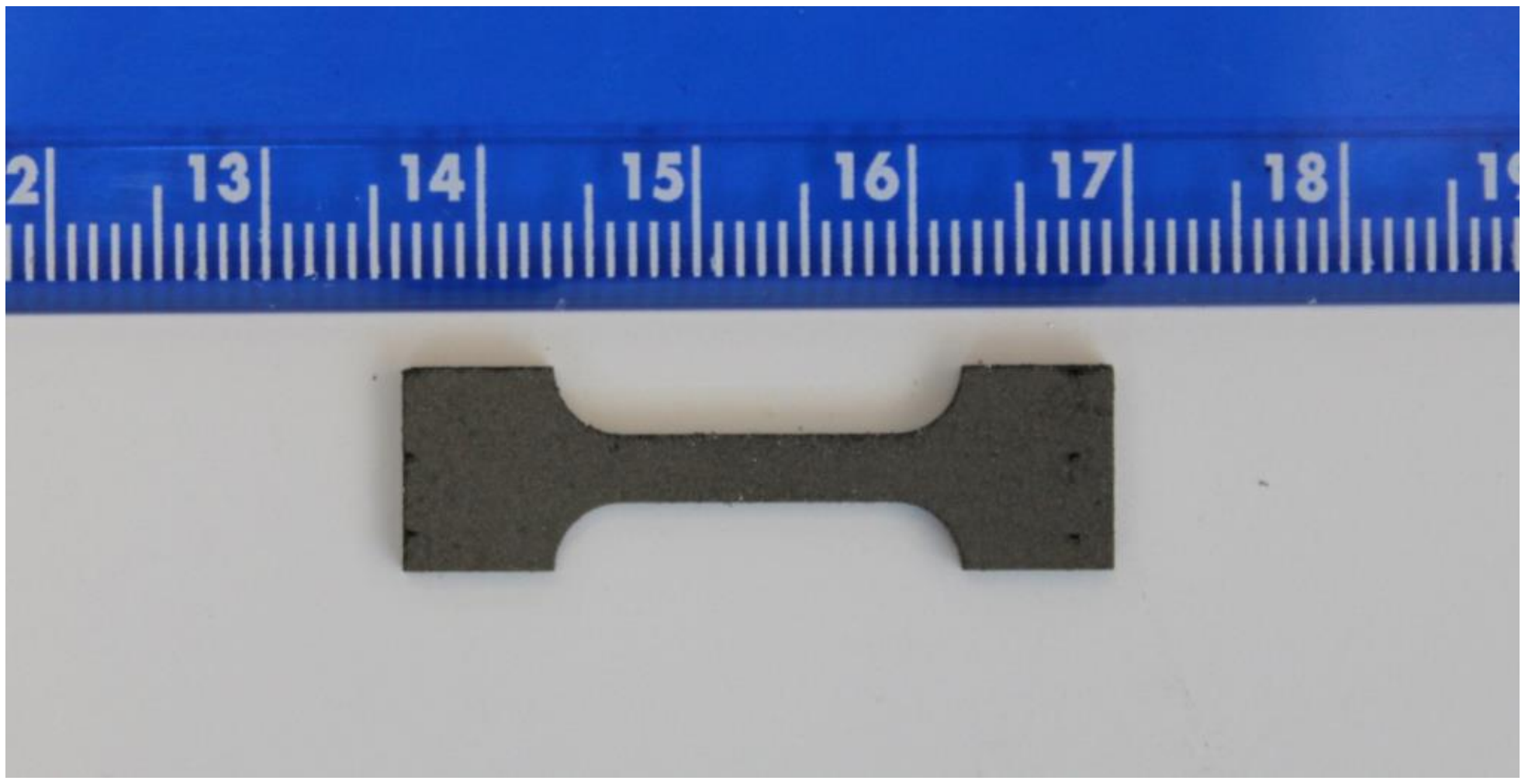

*Fig. S2. Image of a typical dog-bone sample used for tensile tests.*

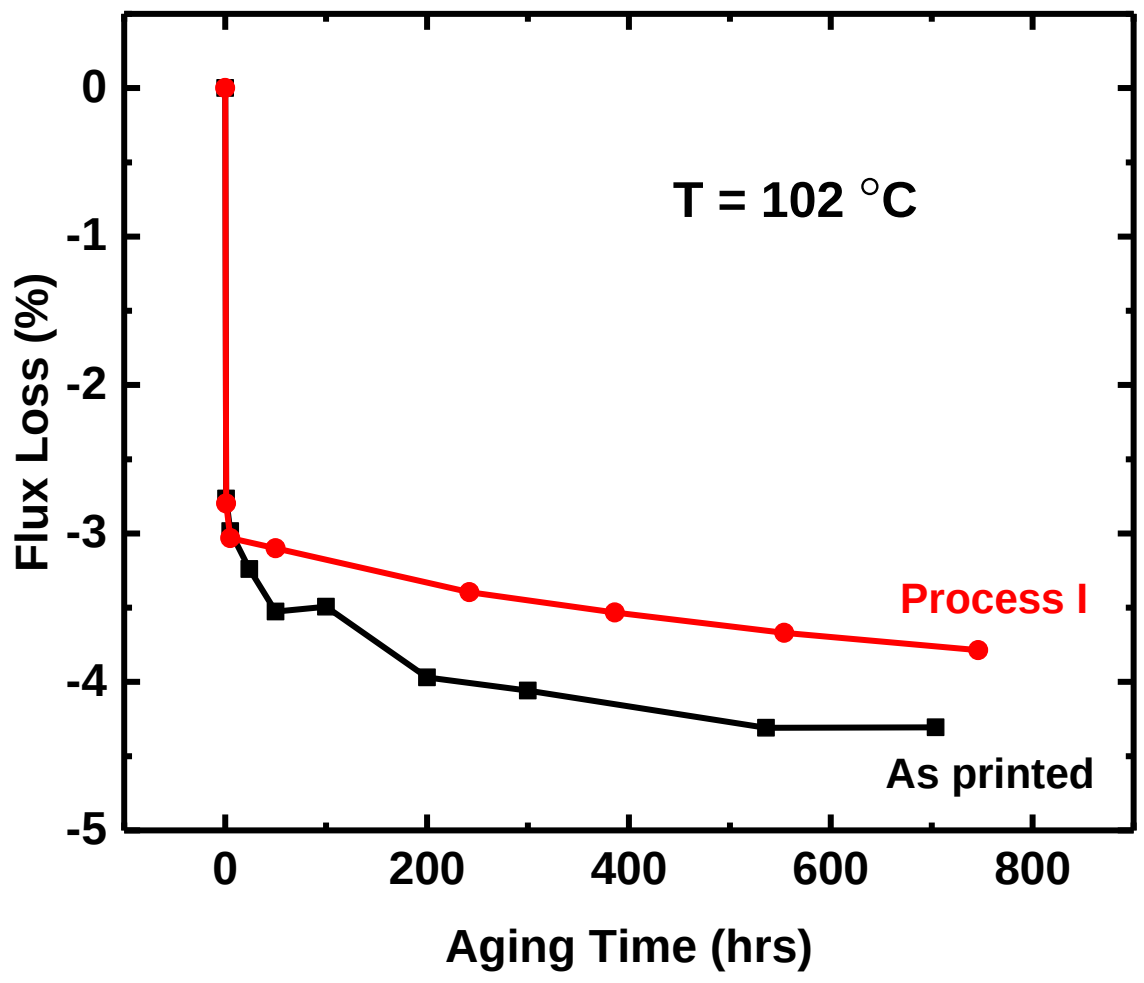

*Fig. S3. Flux aging loss for as-printed and Process I treated 70 vol.% BAAM magnets as a function of aging time at 102 °C*

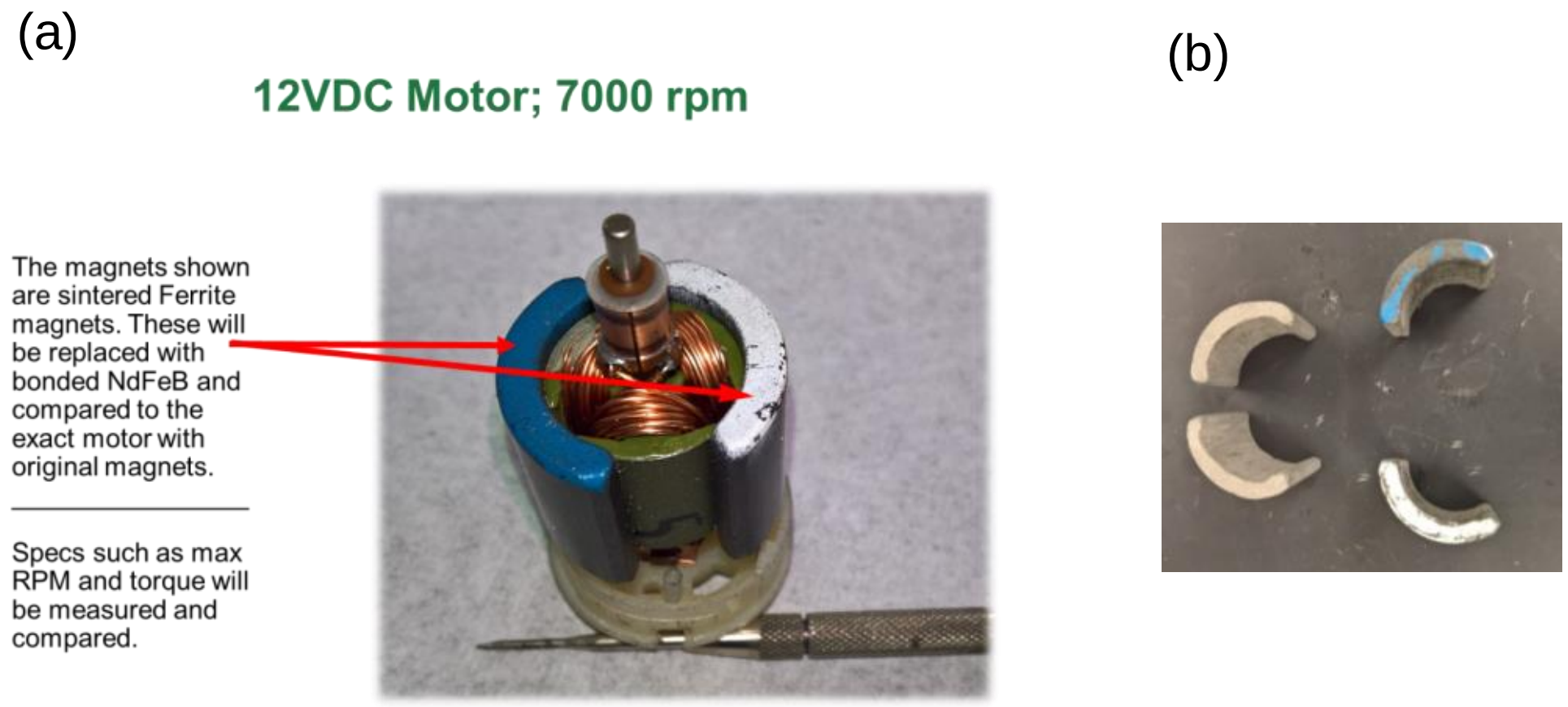

*Fig. S4. (a) Image of the DC motor installed with two arc sintered ferrite magnets; (b) Image of two pairs of sintered ferrite and BAAM 70 vol.% Nd-Fe-B magnets. The arc magnets' dimentions are: height = 26.8 mm, thickness = 4.8 mm, length = 38.2 mm, and angle = 133.5 deg.*

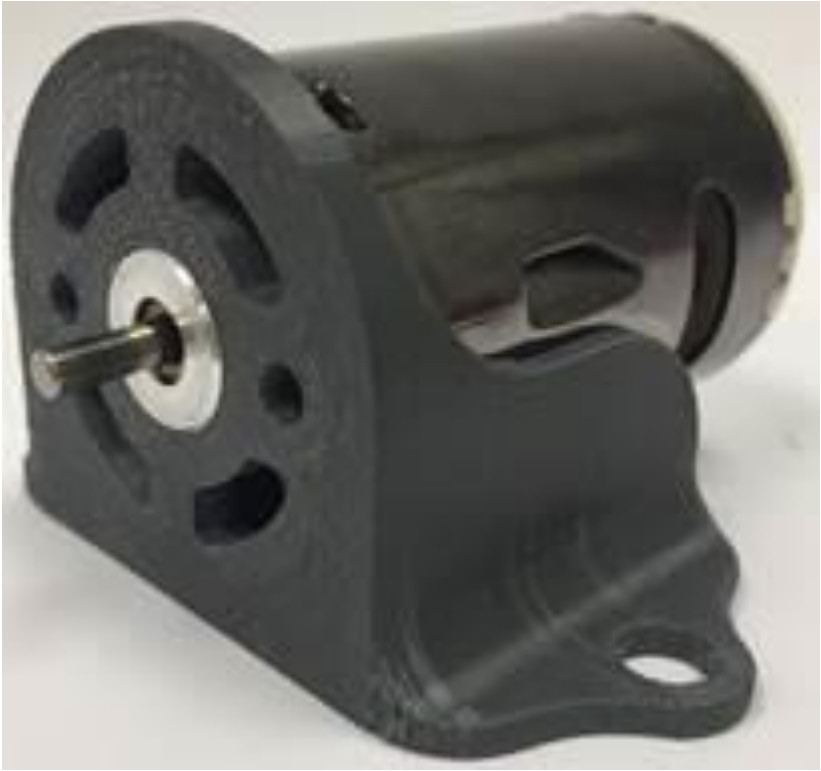

*Fig. S5. DC motor mount.*